\documentclass[twocolumn,authoryear]{aastex63}
\usepackage{graphicx}	
\usepackage{amsmath}	
\usepackage{amssymb}	
\usepackage{natbib}
\usepackage{times}

\usepackage{subfigure}
\subfiguretopcaptrue
\usepackage{url}

\usepackage{etoolbox}

\shortauthors{Zhang et al.}

\begin{document}
\title{Discovery of polarizations from ground state absorption lines: tracer of sub-Gauss magnetic field on 89Her}
\correspondingauthor{Huirong Yan}
\email{huirong.yan@desy.de}

\author[0000-0003-2840-6152]{Heshou Zhang}
\affiliation{Deutsches Elektronen-Synchrotron DESY, Platanenallee 6, D-15738 Zeuthen, Germany}
\affiliation{Institut f$\ddot{u}$r Physik und Astronomie, Universit$\ddot{a}$t Potsdam, Haus 28, Karl-Liebknecht-Str. 24/25, D-14476 Potsdam, Germany}

\author[0000-0002-8364-7795]{Manuele Gangi}
\affiliation{INAF - Osservatorio Astrofisico di Catania, Via S. Sofia 78, I-95123 Catania, Italy}
\affiliation{INAF - Osservatorio Astronomico di Roma, Via Frascati 33, I-00078 Monte Porzio Catone, Italy}

\author[0000-0001-7626-3788]{Francesco Leone}
\affiliation{Universit\'a di Catania, Dipartimento di Fisica e Astronomia, Sezione Astrofisica, Via S. Sofia 78, I-95123 Catania, Italy}

\author{Andrew Taylor}
\affiliation{Deutsches Elektronen-Synchrotron DESY, Platanenallee 6, D-15738 Zeuthen, Germany}

\author[0000-0003-2560-8066]{Huirong Yan}
\affiliation{Deutsches Elektronen-Synchrotron DESY, Platanenallee 6, D-15738 Zeuthen, Germany}
\affiliation{Institut f$\ddot{u}$r Physik und Astronomie, Universit$\ddot{a}$t Potsdam, Haus 28, Karl-Liebknecht-Str. 24/25, D-14476 Potsdam, Germany}

\begin{abstract}
{
{ We report the identification of the polarization of ground state absorption lines from post-AGB 89~Hercules. Two ground state neutral iron lines are found to have counterintuitive high-amplitude polarizations and an unchanged polarization direction through the orbital period, as opposed to the pattern of polarizations of absorption lines from excited states, which are synchronized with the orbital phase owing to optical pumping. This can be explained with magnetic realignment on the ground state. The 3D mean magnetic field is thereby unveiled from the degree and direction of the polarizations of the two iron lines. The field strength is also constrained to be $\lesssim 100$\,mG. Our result has thus improved the accuracy by orders of magnitude compared to the previous $10\,{\rm G}$ upper limit set by non-detection of the Zeeman effect.}
}
\end{abstract}

\section{Introduction}

{ Spectral polarimetry observations can provide exclusive information on magnetic field and topology of the radiation structure. Developments on high-resolution spectral facilities with polarimeters (e.g., PEPSI, \citealt{PEPSIob}; HARPS, \citealt{HARPSob}; HANPO, \citealt{HANPOob}) allows the investigation of polarimetric properties of single spectral lines.
{ The surprisingly high polarimetric signals shown by the prototypical star 89~Herculis (in average, more than $1\%$) give us the unique opportunity to study this system \citep{leone18}.}

89~Herculis ($89~Her$) is a post-AGB binary system. The primary star is an F-type supergiant with a radius $R_{pri}=41 R_{\sun}$ and an effective temperature $T_{eff}=6500 K$ while the secondary is an M-type main sequence with $R_{sec}=0.6 R_{\sun}$ and $T_{eff}=4045 K$. The radius of the secondary orbital track is $r_{orb}=67 R_{\sun}$. $89~Her$ has an orbital period $P_{orb}\sim288$ days and inclination $12^\circ$. It presents a circumstellar environment consisting of two main components: an expanding hour-glass structure and an circumbinary rotating disk \citep{Waters:1993aa,Kipper:2011aa,Bujarrabal07}. The projection of stellar outflows on the picture plane is $45^\circ$ comparing to the East-West orientation \citep{Bujarrabal07}. { \citet{leone18} has found that linearly polarized photospheric absorption lines of 89~Her present $Q/I$ and $U/I$ signals varying according to the secondary orbital period (hereafter {\it ``orbital synchronization''}). They have excluded the origin of such signal from either the depolarization of stellar continuum polarization (incl. pulsations, hot spots), or scattered polarization from a bipolar outflows, and found that it is a result of the optical pumping in the stellar environment.}

In this letter, we report the discovery of two ground state Fe\,{\sc i} photosphere absorption line polarizations, whose directions have not shown { \it ``orbital synchronization''}, but rather are aligned through all orbital phases\footnote{Hereafter, we will assume the same ephemeris adopted in \citep{leone18} to compute the orbital phases of 89\,Her ($iph$).}. This observation indicates the existence of ground state magnetic alignment, which was first theoretically proposed as a magnetic tracer in \citep{YLfine}.} The alignment is in terms of the angular momentum of atoms and ions (hereafter ``atoms'' for simplicity). The ground state alignment (henceforth GSA) is an established physical phenomenon which has solid physical foundations and has been studied and supported by numerous experiments \citep{KASTLER-1950-234250,refId0,Hawkins:1953dz,Hawkins:1955fv,Cohen-Tannoudji:1969fk}.
The anisotropic radiation aligns the atoms on the ground state by optical pumping \citep{Happer:1972ij,Varshalovich:1971mw,Landolfi:1986lh}. These radiative aligned atoms are magnetically realigned by fast precession as long as the Larmor precession rate $\nu_{\rm Lar}$ is larger than the radiative pumping rate $\nu_{\rm Rad}$ \citep{YLfine,YLhyf}. In the GSA regime, the atoms are aligned with radiation field or realigned by magnetic field depending on the ratio of the two rates ($r_{\rm A}\equiv\nu_{\rm Lar}/\nu_{\rm Rad}$). In the Ground level Hanle regime, however, the atoms preferentially follow neither the magnetic or radiation field ($r_{\rm A}\sim 1$) \citep{YLHanle}. The GSA probe is particularly suitable for sub-Gauss magnetic field since the atoms have long life time in their ground states. The resulting absorption lines from the aligned atoms are polarized parallel or perpendicular to the magnetic field direction.

{
The magnetic field in the 89Her's primary has been unknown. Earlier attempts with current magnetic tracers have not provided strong constraint: no continuum polarization is detected (see \citealt{Akras2017}) and previous spectropolarimetric studies have revealed that the source is circularly unpolarized, which only implies an upper limit of 10 G \citep{Sabin15}.
In the letter, we will first discuss the observational results and then provide the theoretical interpretations, which lead to the extraction of the mean magnetic field information on the photosphere of the $89~Her$'s primary.

}

\section{Observational Data}

Reduced spectropolarimetric data of $89~Her$ have been collected at the 3.6-m Canada-France-Hawaii Telescope with the Echelle SpectroPolarimetric Device for the Observation of Stars \citep[ESPaDOnS: R=68000,][]{ESPaDOnSob}. They consist of Stokes $I$, $Q$, $U$, and \emph{null} $NQ$, $NU$ \citep{Leone:2016aa}. From these data we computed the polarization $P$, the \emph{null} polarization $NP$ and the polarization angle $\xi$, according to the definitions given in \citet{Landi04}. The spectral lines studied are unblended with other lines, as shown in Fig.\,1(\,a-\,c)\footnote{Unlike Na D lines and Balmer lines in 89~Her, where the photospheric line is blended with multiple blue-shifted absorption components as well as redshifted strong interstellar absorption (see, e.g., \citealt{Kipper11,Manu19}).}.
To provide an objective criterion for deciding whether a polarimetric signal is detected across the spectral lines, we follow the statistical test of \citet{Donati:1997aa}. Firstly we computed the reduced $\chi^2$ inside and outside the spectral lines for both $P$ and \emph{null} NP profiles. Then we calculated the $\chi^2$ detection probabilities based on the achieved signal-to-noise ratio (S/N), as shown in Table\,1 in Supplementary.
{ We concentrate on the central part of the absorption line (corresponding to the marked green zone in Fig.~\ref{Fig:geometry}d), which is least influenced by the blue and red shifted emission from outflows.}

\begin{figure*}
\begin{center}
\includegraphics[width=0.9\textwidth]{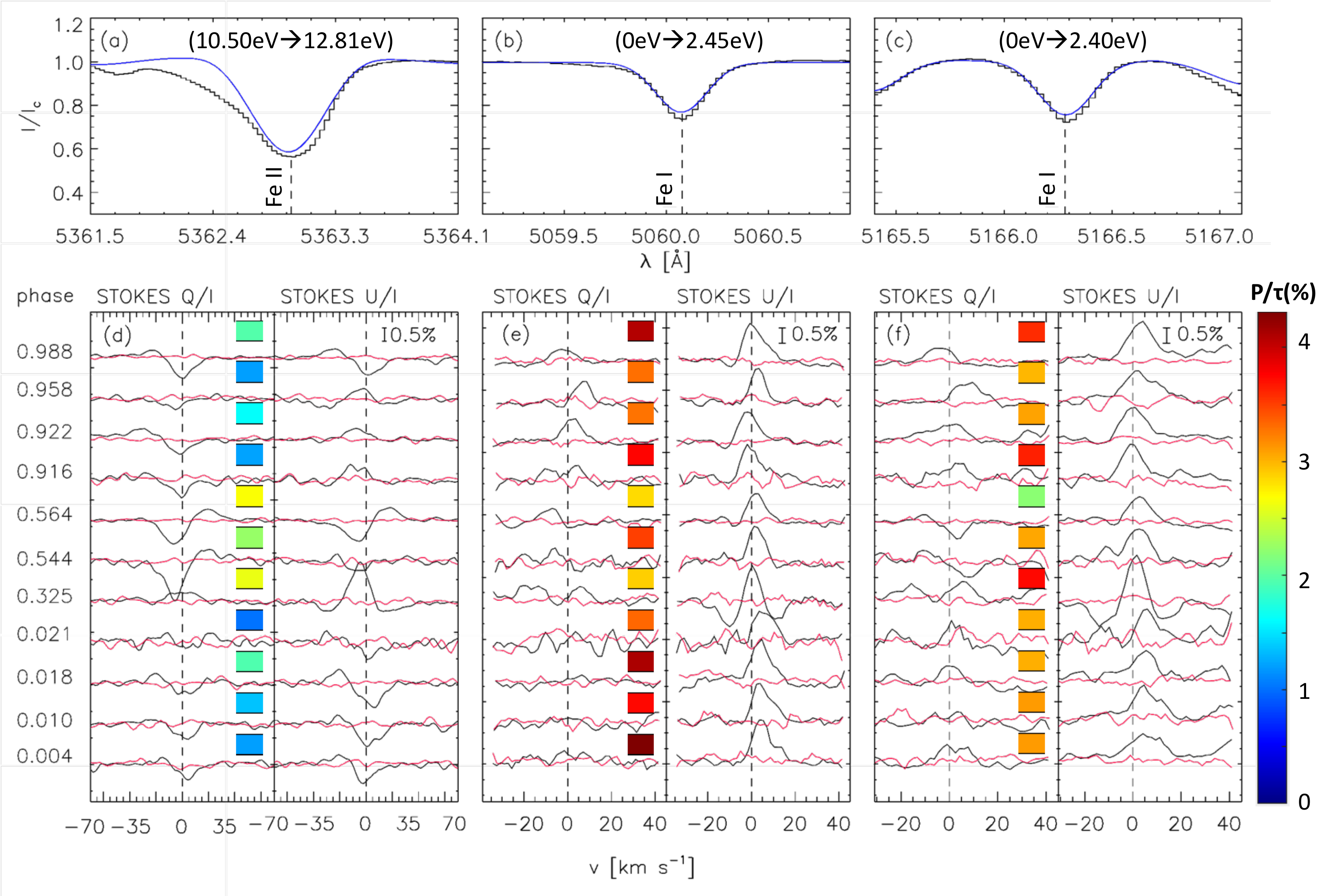}
\end{center}
\caption{{\it Upper:} (a) The observed spectral line profile (black) and its comparison with the photospheric lines modeled (blue) with the synthetic spectrum package SYNTHE \citep{Kurucz2005} for the radiative dominant line Fe\,{\sc ii} $\lambda5362.970$\,\AA\, (a), and of the two magnetic aligned lines  Fe\,{\sc i} $\lambda5060.249$\,\AA\,(b) and Fe\,{\sc i} $\lambda5166.282$\,\AA\, (c). These lines are unblended and identified from photosphere.
{\it Middle:} $Q/I$ and $U/I$ profiles of (d) Fe\,{\sc ii} $\lambda5362.970$\,\AA, (e) Fe\,{\sc i} $\lambda5060.249$\,\AA, (f) Fe\,{\sc i} $\lambda5166.282$\,\AA\, lines. The \emph{null} spectra are marked with red color. The degree of polarization P in each phase is denoted by the color mark. 
}\label{Fig:data}
\end{figure*}


\section{Polarization Analysis for the Photospheric Absorption lines}

Previous spectropolarimetric works focus only on lines with a relatively large optical depth ($\tau \gtrsim 0.5$). { We found that all the photosphere lines showing {\it ``orbital synchronization''} stand for the transitions between the excited states.} As an example, the linear polarization profile of Fe\,{\sc ii} $\lambda5362.970$\,\AA\ ($10.50eV\rightarrow12.81eV$) is presented in Fig.\,1d, showing the correlation between their polarization angle and the vector connecting the two stellar bodies during their orbits (see Fig.\,2a).

Here we focus on two ground state neutral iron absorption lines, Fe\,{\sc i}\,$\lambda5060.249$\,\AA\ ($0eV\rightarrow2.45eV$) and Fe\,{\sc i}\,$\lambda5166.282$\,\AA\ ($0eV\rightarrow2.40eV$). These lines are weak ($\tau \sim 0.2$, see Fig.\,1b,\,c) and therefore were previously ignored. As demonstrated in Fig.\,1(e,\,f), they have shown counterintuitive strong polarization signatures. Moreover, the polarization angles of these two Fe\,{\sc i} lines are aligned through all orbital phases (see Fig.~\ref{Fig:geometry}b) towards $\sim45^\circ$ to East-West orientation.
Such results indicate that the { atomic angular momentum on the ground state is realigned by fast magnetic precession, in other words, the magnetic realignment dominates over optical pumping.}
We note that at $iph=0.325$, the polarization direction of the radiation dominant lines are the same as that of GSA lines. This is because the radiative pumping direction at this phase coincides with the magnetic alignment direction.

We find that only the absorption term of linear polarization is important to account for in the analysis of these two GSA lines. The reasons are the following: (1) The background stellar continuum is unpolarized \citep{Akras2017}; (2) The spontaneous (scattered) emission from ``the upper states''\ are negligible (see Fig.~\ref{Fig:atom}a for microphysics). The upper states of those two transitions have different angular momentum ($J_u=3,5$ respectively). Hence the scattered emission would yield a different polarization profile between the two lines at different $iph$. This is at odds with the observations of the aligned polarization direction.
(3) The polarization signals for absorption from ``the upper states'' also support our conclusion (see Fig.~\ref{Fig:atom}b).
We consider the absorption line Fe\,{\sc i}\,$\lambda4187.039$\,\AA ($2.45eV\rightarrow5.41eV$), i.e., the absorption from ``the upper level'' $2.45eV$. Its Stokes Q and U signals are compared with two types of lines: ``the pumping dominant reference'' Fe\,{\sc ii} $\lambda5362.970$\,\AA\, and the GSA line Fe\,{\sc i}\,$\lambda5060.249$\,\AA. The orbital variation pattern of Fe\,{\sc i}\,$\lambda4187.039$\,\AA ($2.45eV\rightarrow5.41eV$) is similar to the Fe\,{\sc ii} $\lambda5362.970$\,\AA\, line, showing a varying radiative alignment of atoms on $2.45eV$. This level's varying contribution to the unchanged polarization signal of the ground state absorption line Fe\,{\sc ii} $\lambda5060.249$\,\AA\, is therefore negligible. Magnetic alignment happens only on the ground state.

\section{Theoretical Analyses}

\subsection{Theoretical expectations and the identification of the 2D magnetic field}
\label{theory}

We solve the atomic transition equations to estimate the expected level of polarization signals from $89~Her$ (see App. B for details). The ground state is taken to be fully magnetically aligned { and the anisotropic pumping is provided by the secondary}.
The $90^\circ-$ambiguity \citep[known as Van Vleck ambiguity,][]{VV25, House74} between the observed polarization and the position of sky (henceforth POS) magnetic projection is also resolved from the comparison of theoretical expectations and the observed degree of polarization. We scan the full parameter space to solve the expected maximum alignment parameter and the ratio of polarization $P/\tau$. { Given the face-on orbit, we take the radiation from the secondary in the range $\theta_0\in[65^\circ,90^\circ]$. We find that in the case of parallel alignment ($\varepsilon_\sigma=+1$), the expected maximum polarization signal can reach $P_{max}/\tau(5060\AA)\gtrsim 8\%, P_{max}/\tau(5166\AA)\gtrsim4\%$, respectively.} These upper limits for the two lines are consistent with the observed polarization signals of both the ground state absorption lines.
Nonetheless, the maximum $P_{max}/\tau$ in the case of perpendicular alignment ($\varepsilon_\sigma=-1$) calculated from the transitional equation are $4.9\%$ and $2.5\%$ for the two absorption lines $\lambda5060.249$\,\AA, $\lambda5166.249$\,\AA\,, respectively, smaller than the observed values. The perpendicular alignment is thus excluded.

\subsection{Magnetic field in the 3rd dimension}

\begin{figure*}
\begin{center}
\includegraphics[width=1\textwidth]{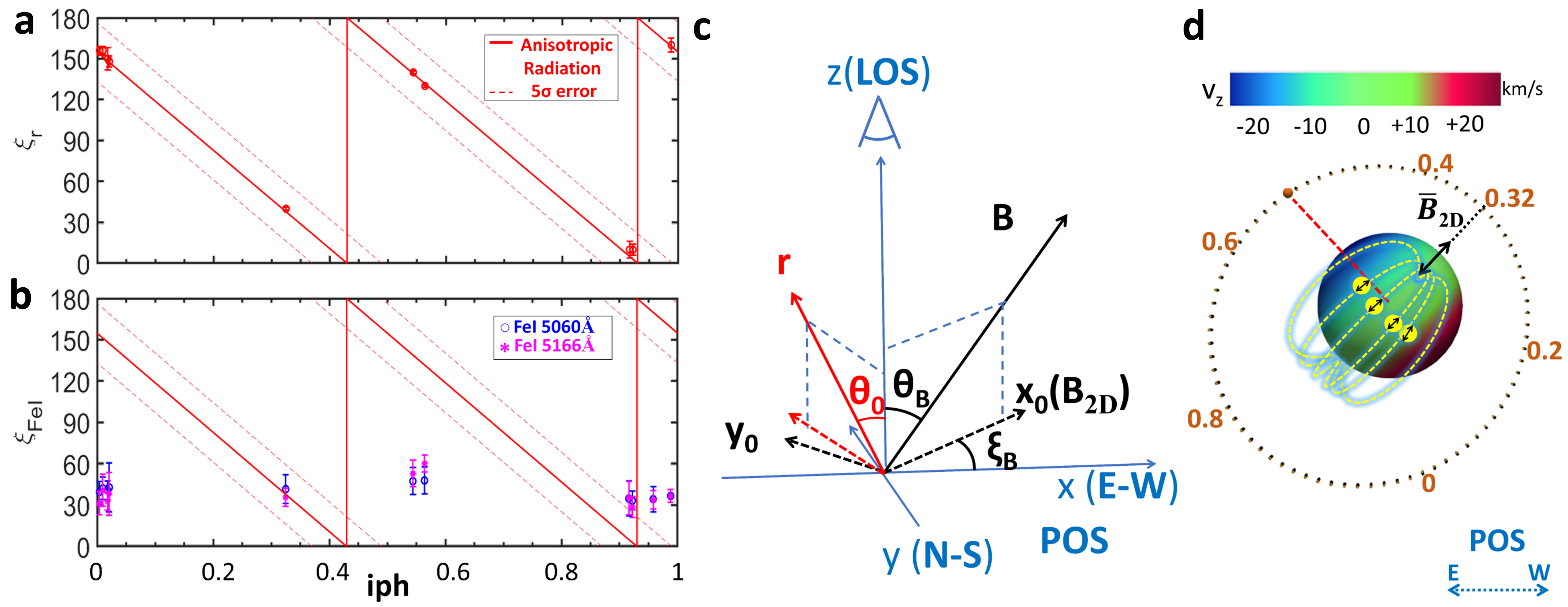}
\end{center}
\caption{(a) strong absorption lines between excited states, the polarization angle ($\xi_{r}$) { shows {\it ``orbital synchronization''}}. $iph$ is orbital phase; (b) the polarization angle $\xi_{\rm FeI}$ for the two Fe\,{\sc i} lines. The error bars mark the $3\sigma$ uncertainty range. (c) 3D view of the system showing the reference frame. $xyz-$frame is the observational frame where $ox-$axis is defined arbitrarily on East-West direction. $\theta_B$ and $\xi_B$ are the polar and azimuth angles for magnetic field in the frame. $x_0y_0z-$frame is the theoretical frame where $ox_0-$axis is the direction of the plane-of-sky magnetic field projection. (d) $89~Her$ system. The orbital phases are marked along the secondary track. { The color scale denotes the line-of-sight (henceforth LOS) velocity ($v_z$) of the photosphere medium. { Our analysis measures the mean magnetic field in the green region of the primary's photosphere, corresponding to the central part of the absorption lines. A possible magnetic tomography of the system is presented here.} The yellow dashed lines are magnetic field lines. The double arrows are the polarization directions. The average magnetic field projection on POS $\bar{B}_{2D}$ is $\sim45^\circ$ to the East-West direction pointing to the orbital phase $0.32$.}}
\label{Fig:geometry}
\end{figure*}

The theoretical Stokes parameters [$\tilde{I}, \tilde{Q}, \tilde{U}, \tilde{V}$] are defined with $\tilde{Q}$ measured from the magnetic field direction on the plane of sky (see $x_0y_0z-$frame Fig. 2c). { In the case of purely absorbing medium}, they are given by \cite{YLfine}:
\begin{equation} \label{eq:Stokes}
\begin{split}
\tilde{I} &= (I_0+Q_0)e^{-\tau(1+\eta_1/\eta_0)} + (I_0-Q_0)e^{-\tau(1-\eta_1/\eta_0)} ,\\
\tilde{Q} &= (I_0+Q_0)e^{-\tau(1+\eta_1/\eta_0)} - (I_0-Q_0)e^{-\tau(1-\eta_1/\eta_0)} ,\\
\tilde{U} &= U_0e^{-\tau} , \tilde{V}=V_0e^{-\tau},
\end{split}
\end{equation}
where $\eta_i$ are absorption coefficients of the Stokes parameters. Theoretically, the absorption coefficients are determined by \citep[see][]{Landi-DeglInnocenti:1984kl}:
\begin{equation} \label{eq:eta}
\begin{split}
\eta_i(\nu, \Omega) &= \frac{h\nu_0}{4\pi} B n(J_l) \xi(\nu-\nu_0)\sum_{kq} (-1)^k \\ &\times w^{(k)}_{J_lJ_u}  \sigma^k_q(J_l) \mathcal{J}^k_q(i, \Omega).
\end{split}
\end{equation}
where $\sigma^k_q(J_l)\equiv\rho^k_q(J_l)/\rho^0_0(J_l)$, $\rho^k_q(J_l)$ is the irreducible density matrix on the ground state. In the GSA regime, only $q=0$ is non-zero and $\sigma^2_0(J_l)$ is the alignment parameter \citep{YLfine}. The quantity $ n(J_l)$ is the population on the lower level. The quantity $ \mathcal{J}^k_q(i, \Omega)$ is the irreducible unit tensors for Stokes parameters I, Q, and U. The ratio $\eta_1/\eta_0$ in the GSA regime is then \citep{YLfine}:
\begin{equation} \label{eq:c}
\eta_1/\eta_0 = -\frac{1.5\sigma^2_0(J_l)\sin^2\theta_B w^{(2)}_{J_lJ_u}}{\sqrt{2}+\sigma^2_0(J_l)(1-1.5\sin^2\theta_B) w^{(2)}_{J_lJ_u}},
\end{equation}
where $\theta_B$ is the angle between magnetic field and the line of sight.
\begin{equation}
w^{(2)}_{J_lJ_u}\equiv\left\{\begin{array}{ccc}1 & 1 & 2\\J_l&J_l& J_u\end{array}\right\}/\left\{\begin{array}{ccc}1 & 1 & 0\\J_l&J_l& J_u\end{array}\right\},\nonumber
\label{w2}
\end{equation} is determined by the electron configurations of the transition $J_l \rightarrow J_u$.
The alignment parameter $\sigma^2_0$ is the same for two transitions with the same ground level ($J_l\rightarrow J_{u1}$, $J_l\rightarrow J_{u2}$). Different observed degree of polarization for these two lines results from the angular momentum configuration parameters $w^{(2)}_{J_{l}J_{u1}}, w^{(2)}_{J_{l}J_{u2}}$. Therefore, we obtained from Eq.(\ref{eq:c}):
\begin{equation} \label{eq:Set}
\begin{split}
&\sigma^2_0(J_l)=\\
&\left\{
             \begin{array}{lr}
             &\frac{\sqrt{2}(c_1c_2(w^{(2)}_{J_{l}J_{u2}}-w^{(2)}_{J_{l}J_{u1}})+\varepsilon_{\sigma}(c_1w^{(2)}_{J_{l}J_{u2}}-c_2w^{(2)}_{J_{l}J_{u1}}))}{(c_2-c_1)w^{(2)}_{J_{l}J_{u1}}w^{(2)}_{J_{l}J_{u2}}},\\ & {\rm sgn}(w^{(2)}_{J_{l}J_{u1}})={\rm sgn}(w^{(2)}_{J_{l}J_{u2}}); \\
             &-\frac{\sqrt{2}(c_1c_2(w^{(2)}_{J_{l}J_{u2}}-w^{(2)}_{J_{l}J_{u1}})+\varepsilon_{\sigma}(c_1w^{(2)}_{J_{l}J_{u2}}+c_2w^{(2)}_{J_{l}J_{u1}}))}{(c_1+c_2)w^{(2)}_{J_{l}J_{u1}}w^{(2)}_{J_{l}J_{u2}}},\\ & {\rm sgn}(w^{(2)}_{J_{l}J_{u1}})=-1, {\rm sgn}(w^{(2)}_{J_{l}J_{u2}})=+1;
             \end{array}
               \right.\\
&\sigma^2_0(J_l)\sin^2\theta_B=\\
&\left\{
             \begin{array}{lr}
             &\frac{2\sqrt{2}c_1c_2(w^{(2)}_{J_{l}J_{u2}}-w^{(2)}_{J_{l}J_{u1}})}{3(c_2-c_1)w^{(2)}_{J_{l}J_{u1}}w^{(2)}_{J_{l}J_{u2}}},\\
             & {\rm sgn}(w^{(2)}_{J_{l}J_{u1}})={\rm sgn}(w^{(2)}_{J_{l}J_{u2}}); \\
             &-\frac{2c_1c_2(w^{(2)}_{J_{l}J_{u2}}-w^{(2)}_{J_{l}J_{u1}})}{3(c_1+c_2)w^{(2)}_{J_{l}J_{u1}}w^{(2)}_{J_{l}J_{u2}}},\\
             & {\rm sgn}(w^{(2)}_{J_{l}J_{u1}})=-1, {\rm sgn}(w^{(2)}_{J_{l}J_{u2}})=+1;
             \end{array}
               \right.
\end{split}
\end{equation}
The ratio $c_{j}\equiv |\eta_1/\eta_0|=\frac{1}{2\tau_j}\ln\left(\frac{1+P_{j}}{1-P_j}\right),\,j=1,2$ is directly related to the degrees of polarization and optical depths of a doublet (see Eq.~\ref{eq:Stokes}). The quantity $\varepsilon_{\sigma}$ is the sign of the alignment parameter for the lower level $\sigma^2_0(J_l)$. The positive alignment $\varepsilon_{\sigma}=+1$ corresponds to the alignment of the angular momentum on ground states being parallel to the magnetic field, whereas the negative alignment $\varepsilon_{\sigma}=-1$ means it is perpendicular. { The averaged magnetic field polar angle in the absorbing volume can hence be achieved by $\overline{\sin^2\theta_B}={\int dV_{abs}\sigma^2_0(J_l)\sin^2\theta_B}/{\int{dV_{abs}\sigma^2_0(J_l)}}$.}

In the case of the two Fe\,{\sc i} absorption lines, the configuration parameters of these two transitions are of the same sign: $w^{(2)}_{J_{l}J_{u1}}=\omega^{(2)}_{4,3}=0.4432; w^{(2)}_{J_{l}J_{u2}}=\omega^{(2)}_{4,5}=0.2256$. Therefore, the corresponding results are:
\begin{equation} \label{eq:FeI_ana}
\begin{split}
\sigma^2_0(J_l)&=\frac{-0.2176c_1c_2+\varepsilon_{\sigma}(0.2256c_1-0.4432c_2)}{0.0707(c_2-c_1)},
\end{split}
\end{equation}

\begin{figure*}[ht]
\begin{center}
\includegraphics[width=1\textwidth]{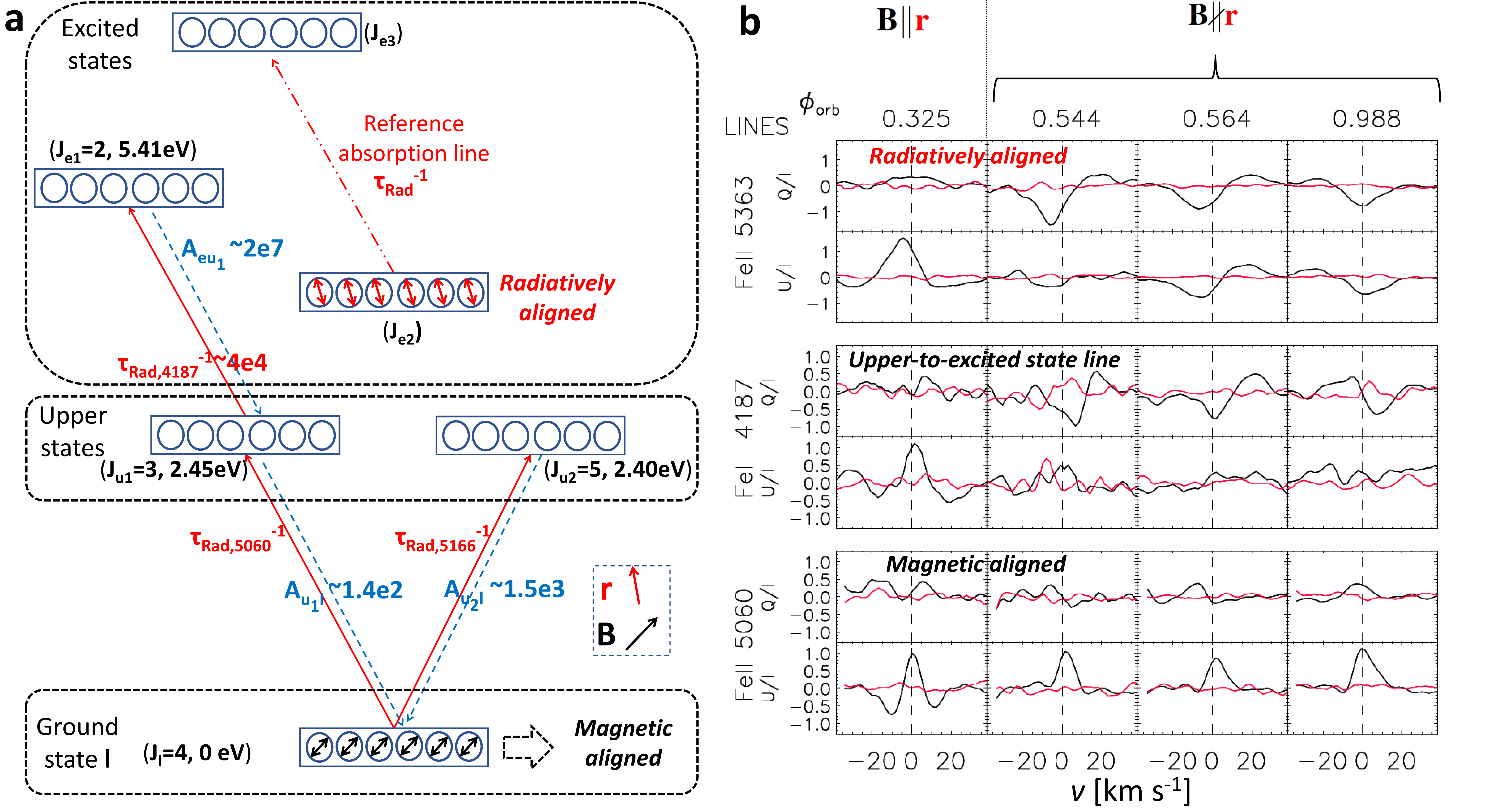}
\end{center}
\caption{(a) Schematics of the optical pumping and magnetic realignment in the GSA regime. Rectangles with circles represent the energy state occupied by atoms. The arrows are the atomic angular momentum. The ground state is magnetically aligned. ``The upper states'' mark the upper levels for transitions of focus Fe\,{\sc i}\,$\lambda\lambda5060, 5166$\,\AA. Other states are marked as excited states. The Einstein coefficients for transitions involving the level Fe\,{\sc i} $2.45eV$ are marked. (b) Stokes parameters of the selected absorption lines. They are: the pumping reference Fe\,{\sc ii} $\lambda5362.970$\,\AA; Fe\,{\sc i}\,$\lambda4187.039$\,\AA\ from ``the upper state'' $2.45eV$; the GSA line Fe\,{\sc i}\,$\lambda5060.249$\,\AA\, at 4 different orbital phases: 0.325, 0.544, 0.564, 0.988.}\label{Fig:atom}
\end{figure*}

The alignment is parallel, i.e., $\varepsilon_\sigma=1$. $\theta_B$ can then be also obtained by solving the Eq.(\ref{eq:FeI_ana}). The resulting $\theta_B, \xi_B$ and their error bars in different phases are presented in Table 1 { (see Fig.~\ref{Fig:geometry}c for 3D geometry)}. { As illustrated in  Fig.~\ref{Fig:geometry}d, a dipole magnetic field on the stellar surface can support such analysis reasonably, with the average 2D magnetic field parallel to the outflow orientation on POS.}

\begin{table}[ht]
\caption{\label{tab:3D} 3D magnetic field angles. The POS component $\xi_B$ and LOS component $\theta_B$ as demonstrated in Fig. 2c. $\Delta\xi_B$ is the 3$\sigma$ error bar obtained from the observed direction of polarization and $\Delta\theta_B$ is the 1$\sigma$ error bar inferred from the degree of the polarization (\S4.1).}
\begin{center}
\begin{tabular}{ccccc}
\hline
\multicolumn{1}{c}{iph}   & \multicolumn{1}{c}{$\xi_B$} & \multicolumn{1}{c}{$\Delta\xi_B$}  & \multicolumn{1}{c}{$\theta_B$} & \multicolumn{1}{c}{$\Delta\theta_B$} \\
\hline
0.004	& 35.5	& 7.9	& 62.4	& 11.0 \\
0.01	& 41.5	& 10.0	& 56.8	& 15.9 \\
0.018	& 36.8	& 6.6	& 58.6	& 13.6 \\
0.021	& 40.5	& 16.6	& 54.9	& 14.6 \\
0.325	& 38.7	& 8.5	& 55.6	& 11.9 \\
0.544	& 50.2	& 9.7	& 59.3	& 11.1 \\
0.564	& 54.0	& 8.0	& 53.2	& 11.0 \\
0.916	& 35.0	& 12.3	& 57.0	& 15.6 \\
0.922	& 30.9	& 7.2	& 54.5	& 13.8 \\
0.958	& 34.1	& 8.1	& 56.0	& 14.3 \\
0.988	& 36.7	& 4.7	& 67.6	& 8.3 \\
\hline
\end{tabular}
\end{center}
\end{table}

\begin{figure}
\begin{center}
\includegraphics[width=1\columnwidth]{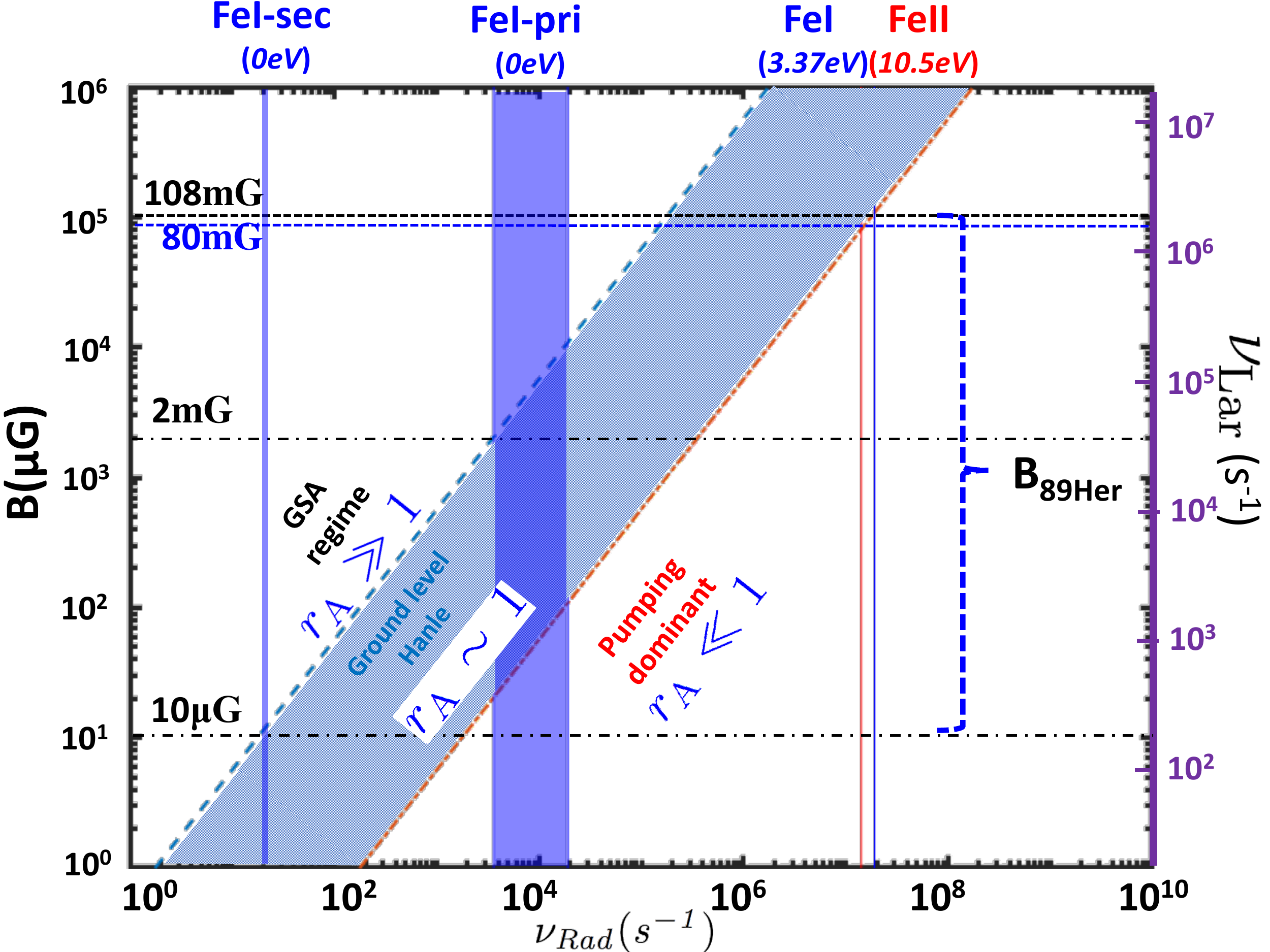}
\end{center}
\caption{Magnetic field strength in 89~Her. The escaping rates from Fe\,{\sc i}($0eV$), and from Fe\,{\sc i}($3.37eV$), Fe\,{\sc ii}($10.5eV$), are marked on top. The dash-dotted lines mark the lower limits of magnetic field strength, { estimated based on the assumption of optical pumping by the secondary and primary, respectively}. The dashed lines show the upper limit of magnetic field strength derived from different transitions.
}\label{Fig:strength}
\end{figure}

\subsection{Constraining the magnetic field strength}

{ Magnetic field strength of $89~Her$ is not directly available since both the radiative alignment on the upper states and the magnetic realignment are saturated effects. Nonetheless, the magnetic field strength can be constrained by comparing the life time of different states (either radiative pumping rate or spontaneous emission rate) with the Larmor precession rate ($\omega_{Lar} \sim17.6({\rm B}/\mu{\rm G})$).}
On one hand, the radiative pumping rate from the Fe\,{\sc i} ground state\footnote{All transitions from the given level are taken into account (both excitations and spontaneous emission) when the escaping rate from a level is calculated. The NIST Atomic Spectra Database is adopted for the Einstein coefficients to calculate the pumping rate. A total of $47$ transitions from Fe\,{\sc i} ground state and $38$ transitions from the state Fe\,{\sc ii}($10.50eV$) are included.} is much smaller than the Larmor precession rate, ${\it max}\{\nu_{Rad}\}<0.1\nu_{Lar}$.
With the secondary providing the optical pumping, the magnetic field strength lower limit is $\sim10{\mu}G$ (see Fig.~\ref{Fig:strength}).

On the other hand, the synchronization between pumping direction and polarizations of lines from the excited states shows that the escaping rate from the excited states, which includes both radiative pumping and spontaneous emission, dominates over the Larmor precession rate on the excited states. Hence the upper limit of the magnetic field strength can be narrowed by $10\,\nu_{Lar}<{\it max}\{\nu_{Rad}, \nu_{Em}\}$.
Utilizing the fact that Fe\,{\sc ii}\,$\lambda5362.970$\,\AA\,($10.50eV\rightarrow12.81eV$) absorption line is fully radiative aligned for all the orbital phases, we deduce the upper limit field strength being $\sim108{\it mG}$. Additionally, we find that the Fe\,{\sc i}\,$\lambda5586$\,\AA\,($3.37eV\rightarrow5.59eV$) absorption line presents a more than $3\sigma$ detectability polarization signal at the orbital phases $0.544, 0.564, 0.988$ and that the polarization direction for all those phases aligns with radiative pumping direction rather than magnetic field direction. This gives us an upper limit for the field strength of $\sim80\,{\it mG}$. The accuracy of the field strength is increased by at least two decades compared to the previous $10\,{\rm G}$ upper limit, constrained from the non-detection of the Zeeman effect.

\section{Discussion}

{ GSA is the most probable and natural explanation for the discovered polarization signals of the ground state absorption lines, that are realigned to one direction as opposed to the { \it ``orbital synchronization''}. We provide here the mean magnetic field of the photosphere surface medium { that corresponds to the central part of the absorption line. Fig.~\ref{Fig:geometry}d serves illustratively that such magnetic field is reasonable on the stellar surface.} A full magnetic field tomography requires further studies, which is beyond the scope of current observation precision. Higher resolution spectral observation will enable us to study the wings in the absorption line profiles, separating the emission components from outflows.

Through our analyses, we assume the anisotropic pumping comes from the secondary (following \citealt{leone18}). However, when constraining the magnetic field strength, a narrower range might be achieved if accounting for the photons from the primary. We calculate the extended radiation field of the primary following \citep{zhang15} with the Lambert cosine law and limb-darkening \citep{CB11} considered, and find that the lower limit for magnetic field strength can be $\sim 2 mG$ (see Fig.~\ref{Fig:strength}a). { Additionally, we could interpret the {\it ``orbital synchronization''} as a result of local polarized flux illuminating the medium rather than pumping from the secondary. Nonetheless, even under such assumption, the realigned polarization direction of two ground state lines we report here can still be explained by the magnetic realignment of the angular momentum on the ground state.}

Moreover, beyond optical band, GSA can be implemented with multi-frequency data ranging from UV to submillimeter to trace not only the spatial but also the temporal variations of magnetic field \citep{YLhyf,YL12,Shangguan2013,ZY18}.

\section{Summary}
Polarizations of absorption lines from ground state are discovered with high polarization degree for the first time on the binary system of $89~Her$. We conclude that:
\begin{itemize}
\item The polarization directions of the two Fe I lines show little variation across the orbital period as opposed to the polarization of many other absorption lines from the upper states in the same environment, indicating that the atoms are realigned by magnetic field on the ground state.
\item The direction of polarization in the GSA regime directly points to the 2D projection of the mean magnetic field in the photosphere of the primary.
\item The 90 degree (Van Vleck) degeneracy is broken from the comparison of theoretical expectation and the observed degree of polarizations.
\item The polar angle between magnetic field and LOS is obtained from the analysis of the polarization degree of both Fe I lines.
\item The upper limit of the mean magnetic field is $\lesssim 100$mG in the photosphere of the primary from the fact that the polarization from upper levels absorption lines are aligned by the radiation.
\end{itemize}}

\acknowledgements
{We acknowledge helpful communications on various aspects of the paper with the following colleagues: F. Boulanger, S. Gao, J. Liu, R. Liu, K. Makwana, Q. Zhu. MG acknowledges the partial support from DESY during his visit there.}

\bibliography{yan}
\bibliographystyle{aasjournal}

\end{document}